*Article*

# Optical Harmonic Vernier Effect: A New Tool for High Performance Interferometric Fibre Sensors

**André D. Gomes [1,2,*], Marta S. Ferreira [1,3], Jörg Bierlich [1], Jens Kobelke [1], Manfred Rothhardt [1], Hartmut Bartelt [1] and Orlando Frazão [2]**

[1] Leibniz Institute of Photonic Technology (Leibniz-IPHT), Albert-Einstein-Strasse 9, 07745 Jena, Germany; joerg.bierlich@leibniz-ipht.de (J.B.); jens.kobelke@leibniz-ipht.de (J.K.); manfred.rothhardt@leibniz-ipht.de (M.R.); hartmut.bartelt@leibniz-ipht.de (H.B.)

[2] INESC TEC and Department of Physics and Astronomy, Faculty of Sciences, University of Porto, Rua do Campo Alegre 687, 4169-007 Porto, Portugal; ofrazao@inesctec.pt

[3] i3N and Department of Physics, University of Aveiro, Campus Universitário de Santiago, 3810-193 Aveiro, Portugal; marta.ferreira@ua.pt

* Correspondence: andre.gomes@leibniz-ipht.de;

**Abstract:** The optical Vernier effect magnifies the sensing capabilities of an interferometer, allowing for unprecedented sensitivities and resolutions to be achieved. Just like a caliper uses two different scales to achieve higher resolution measurements, the optical Vernier effect is based on the overlap in the responses of two interferometers with slightly detuned interference signals. Here, we present a novel approach in detail, which introduces optical harmonics to the Vernier effect through Fabry–Perot interferometers, where the two interferometers can have very different frequencies in the interferometric pattern. We demonstrate not only a considerable enhancement compared to current methods, but also better control of the sensitivity magnification factor, which scales up with the order of the harmonics, allowing us to surpass the limits of the conventional Vernier effect as used today. In addition, this novel concept opens also new ways of dimensioning the sensing structures, together with improved fabrication tolerances.

**Keywords:** optical fiber sensor; Vernier effect; Fabry–Perot interferometer

## 1. Introduction

The fast development in many research fields utilizing optical fibers along with the specific technical challenges in their use places strong pressure and new challenges for current optical fiber sensing research. There is an increasing need for sensing structures able to achieve higher sensitivities and resolutions than what conventional fiber sensors can offer. As such, researchers are driven to find new solutions for improved fiber sensors. Recently, in one such improvement, the Vernier effect was applied to fiber sensors. This effect, known for many years due to its application in calipers, consists of two measurement scales with slightly different periods so that the overlap of both improves measurement accuracy [1,2]. Similarly, the optical Vernier effect makes use of two interferometers with slightly shifted interferometric frequencies, normally arranged in series (cascaded configuration) [3]. Ideally, one interferometer is used as a sensor while the other acts as a stable reference. However, in this type of configuration, maintaining one interferometer as a reference can be difficult since usually both interferometers are located physically close to one another. This problem was recently solved by using a 3dB fiber coupler to physically separate the two interferometers into a parallel configuration, preserving the properties of the Vernier effect [4]. However, this work still relies only on the standard Vernier effect. In both configurations, the superposition of the responses from the two interferometers produces a beating pattern containing a large envelope that provides a spectral shift magnification compared to the normal sensing interferometer, allowing higher sensitivities and resolutions to be achieved [5].

The optical Vernier effect applied to optical fiber Fabry–Perot interferometers (FPIs) was first reported in 2014 [6]. Since then, different structures have been proposed for a diverse range of sensing applications [3,7–9]. In fact, this subject has quickly become a hot topic in fiber sensing, with more than two-thirds of the reports published since last year [1–31]. A wide variety of FPI configurations, previously demonstrated in the literature, can be further improved through combination with the Vernier effect, resulting in considerably enhanced sensitivities.

In this paper, to the best of our knowledge, we introduce for the first time an extended concept of an optical harmonic Vernier effect for Fabry–Perot interferometers, illustrated in Figure 1. In this case, the optical path length of one of the interferometers is increased by a multiple (*i*-times) of the optical path length of the second interferometer, plus a detuning factor. The magnification factor provided by the Vernier envelope is enhanced in proportion to the order of the harmonic, allowing unprecedented sensitivities and resolutions to be achieved. Moreover, the effect generates internal envelopes distinct from the upper envelope, which is the only envelope monitored in the fundamental optical Vernier effect. These internal envelopes scale up in number and in free spectral range in proportion to the order of the harmonics, making it easier to track the spectral shift by monitoring the intersections provided by them. This report describes the properties of the optical harmonic Vernier effect, together with an experimental demonstration of the effect. This approach opens new possibilities for the development of novel optical fiber sensors with even higher performances, allowing better control and tuning of the effect dependent on the target application.

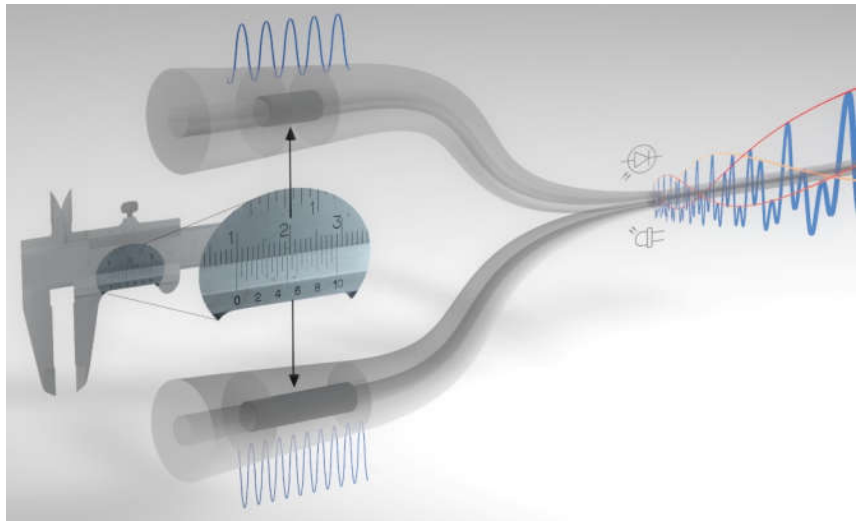

**Figure 1.** Illustration of the harmonic Vernier effect. The Vernier effect, like in a caliper, uses two different scales to achieve higher resolution measurements. Similarly, the optical Vernier effect uses the overlap response of two interferometers with slightly different frequencies. The novel concept of harmonics of the Vernier effect shows that it is, in fact, possible to use two interferometers with very different frequencies, creating a complex harmonic response with enhanced sensing resolution and sensing magnification capabilities when compared to the fundamental optical Vernier effect.

## 2. Theoretical Considerations

### 2.1. Fundamental Optical Vernier Effect

The fundamental optical Vernier effect requires two interferometers with slightly shifted interferometric frequencies. In Fabry–Perot interferometers (FPIs), the interferometric frequency is adjusted by modifying the optical path length of the interferometer. This is achieved by changing the refractive index and/or the physical length of the interferometer. Therefore, given the properties of an initial FPI, the second interferometer can be adjusted to maximize the enhancement provided by the Vernier effect.

In the following analysis, we rely on a parallel configuration using a 3dB fiber coupler [4], where each arm contains a single FPI. This configuration allows both FPIs to be independent of each other, where one of them can easily be maintained as a reference. Note that FPIs positioned in series (without a physical separation provided by an optical coupler) would show equivalent results. However, additional factors would have to be considered in order to describe the effect under such conditions. Although the following theoretical considerations are valid for any FPI structure, they can easily be extended to other types of interferometers, such as the Mach–Zehnder interferometer or the Michelson interferometer, expanding the range of configurations and applications of this powerful technique. Here, we assume that each FPI is an air cavity formed by a silica tube between two sections of single-mode fiber. In this case, all interfaces provide a silica/air Fresnel reflection with an intensity reflection coefficient $R_i$. The value of $R_i$ is small (around 3.3% at 1550 nm), and hence only one reflection at each interface is considered (two-wave approximation). This also allows us to simplify the mathematical description of the effect. Note that we are considering the case of a coherent light source. Therefore, to take into account the phase, we need to make the description in terms of amplitude, and not intensity. In this configuration, the output electric field at the detection system is described as the sum of the electric fields reflected from each interferometer. The output electric field, $E_{out}(\lambda)$, can, therefore, be expressed as (further details in Appendix A):

$$E_{out}(\lambda) = \sqrt{2}AE_{in}(\lambda) + B\frac{E_{in}(\lambda)}{\sqrt{2}}\left\{\exp\left[-j\left(4\pi n_1 L_1/\lambda - \pi\right)\right] + \exp\left[-j\left(4\pi n_2 L_2/\lambda - \pi\right)\right]\right\}, \tag{1}$$

where, $E_{in}(\lambda)$ is the electric field of the input light, $n_1, L_1$ and $n_2, L_2$ are the effective refractive indices and lengths of the first and second interferometer, respectively, and $\lambda$ is the vacuum wavelength. The coefficients $A$ and $B$ are given by:

$$A = \sqrt{R_1}, \tag{2}$$

$$B = (1 - A_1)(1 - R_1)\sqrt{R_2}, \tag{3}$$

where $R_1$ and $R_2$ represent the intensity reflectivities at the cavity interfaces, considered to be similar in both FPIs, $A_1$ represents the transmission losses through the first interface and is related with mode mismatch and surface imperfections. In this approach, no propagation losses are considered. The output reflected light intensity, $I_{out}(\lambda)$, normalized to the input, is then given by (see Appendix A):

$$I_{out}(\lambda) = I_0 - 2AB\left[\cos\left(4\pi n_1 L_1/\lambda\right) + \cos\left(4\pi n_2 L_2/\lambda\right)\right] + B^2\cos\left[4\pi\left(n_1 L_1 - n_2 L_2\right)/\lambda\right], \tag{4}$$

where $I_0 = 2A^2 + B^2$. It shows the combination of the oscillatory responses of both FPIs, plus a lower frequency component given by the difference between the optical path lengths of the two interferometers.

The interference spectrum is modulated by an envelope whose free spectral range (*FSR*) can be described by the relationship between the *FSR*s of each individual interferometer, such as (see Appendix B) [6]:

$$FSR_{envelope} = \left|\frac{FSR_1 FSR_2}{FSR_1 - FSR_2}\right| = \left|\frac{\lambda_1 \lambda_2}{2(n_1 L_1 - n_2 L_2)}\right|, \tag{5}$$

where $\lambda_1$ and $\lambda_2$ are the wavelengths of two adjacent maxima (or minima).

An important characteristic of the Vernier effect is the magnification factor $(M)$. There are currently two definitions for this parameter [17]. In the first, the $M$-factor expresses how large the $FSR$ of the envelope is when compared with the individual sensing FPI, here defined as (see Appendix B) [3]:

$$M = \frac{FSR_{envelope}}{FSR_1} = \left| \frac{n_1 L_1}{n_1 L_1 - n_2 L_2} \right|,$$  (6)

From Equations (5) and (6), it is noticeable that both the $FSR$ of the envelope and the $M$-factor depend on the optical path lengths of the interferometers that form the optical structure.

The second definition of the $M$-factor is directly related to the sensing application, describing how much the wavelength shift of the envelope is magnified in comparison to the wavelength shift of the individual sensing FPI, under the effect of a certain measurand. In this case, the $M$-factor is expressed as:

$$M = \frac{S_{envelope}}{S_{FPI1}},$$  (7)

where $S_{envelope}$ is the sensitivity of the envelope and $S_{FPI1}$ is the sensitivity of the individual sensing FPI, if the second interferometer acts as a reference.

### 2.2. Optical Harmonic Vernier Effect

Harmonics of the optical Vernier effect are provided when the optical path length (OPL) of one of the interferometers is increased by a multiple ($i$-times) of the OPL of the second interferometer. The fundamental Vernier effect relies on the fabrication of two sensors with close OPLs. From a practical point of view, and considering the current fabrication processes of inline FPIs, usually at sub-millimeter scale, this requirement can be challenging and unfeasible in certain situations. On the other hand, it is possible to tailor the properties of the sensor to produce optical harmonics of the Vernier effect, significantly increasing the design possibilities of the sensor. To explore the harmonic properties, let us assume that the OPL of the second interferometer is increased by $i$-times the OPL of the first interferometer $\left( OPL_2 = n_2 L_2 + i n_1 L_1 \right)$. $i$ indicates the order of the harmonic, where for the case of $i = 0$ one ends up in the fundamental optical Vernier effect. The $FSR$ of the second interferometer, depending on the harmonic order, is now defined as:

$$FSR_2^i = \left| \frac{\lambda_1 \lambda_2}{2(n_2 L_2 + i n_1 L_1)} \right|, i = 0, 1, 2 \ldots$$  (8)

Figure 2 presents the numerical simulations of Equation (4) for the fundamental Vernier effect and for the first three harmonic orders, together with the correspondent fast Fourier transform (FFT). The ideal case of no transmission loss related to mode mismatch and surface imperfections, and no propagation losses were considered. The coefficient $A$ was assumed to be 0.04, $B$ as $0.96A$, and $n_1$ and $n_2$ equal to 1 (air). A length of 41 μm was considered for the first interferometer, and a length of 32 μm plus multiples of the first interferometer length ($i \times 41$ μm), depending on the order of the harmonic, was used for the second interferometer. The slight detuning between the OPLs of the two interferometers is visible in the FFT for the fundamental case, where the peaks corresponding to the frequencies of the interferometers are slightly separated.

The harmonics of the Vernier effect regenerate the upper envelope with the same frequency and the same $FSR$ as in the fundamental case, as one can observe in Figure 2 (note that the upper envelope, indicated by a dashed line, was shifted upwards to distinguish it from the internal envelopes). In fact, the $FSR$ of the upper envelope, described by Equation (5), can be rewritten in a more general way (see Appendix C), for any harmonic order, in the form of:

$$FSR_{envelope}^{i} = \left| \frac{FSR_1 FSR_2^i}{FSR_1 - (i+1)FSR_2^i} \right|. \tag{9}$$

This general equation represents the regeneration property of the upper envelope since it turns out to be independent of the order of the harmonic. Moreover, it is interesting to observe that for odd harmonic orders, the upper envelope suffers a π-shift.

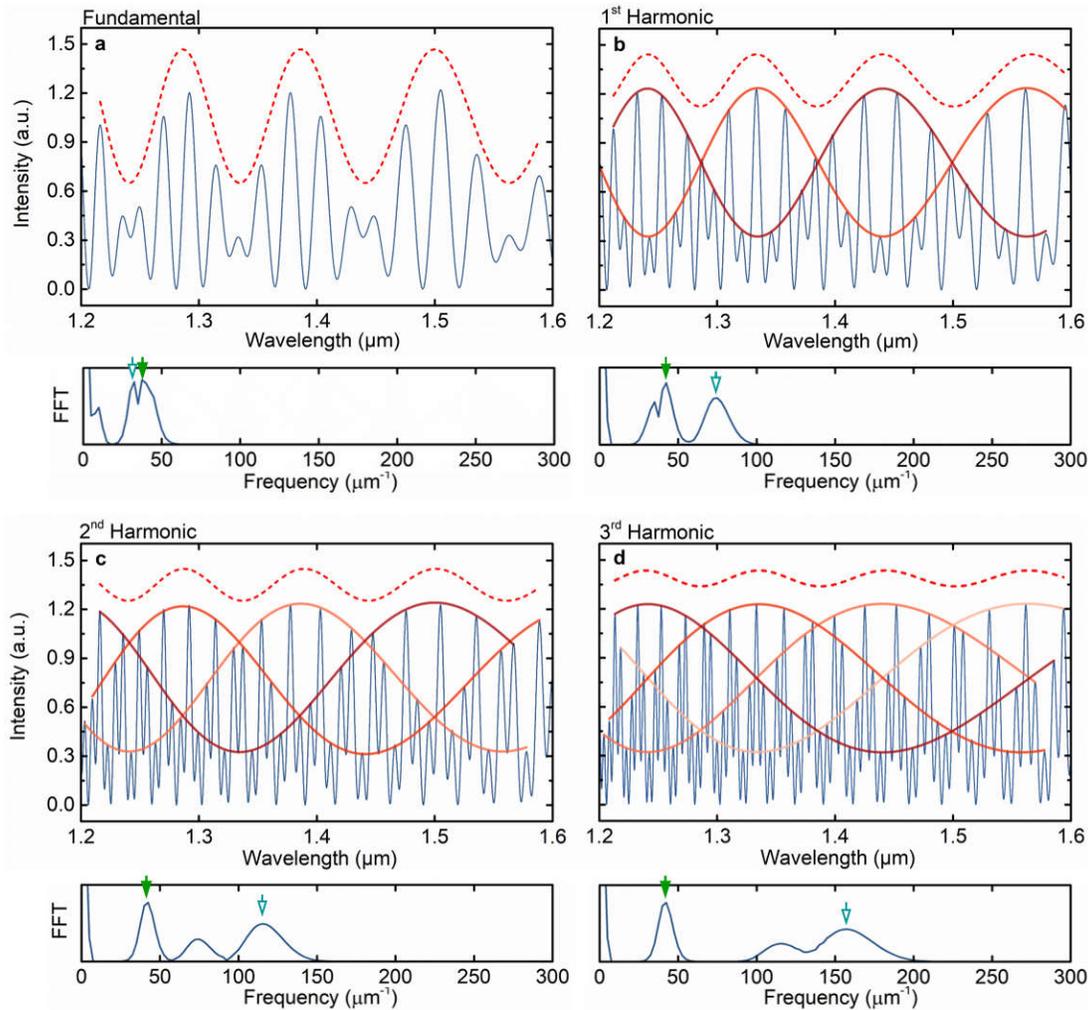

**Figure 2.** Reflected intensity spectra described by Equation (4) in four different situations and corresponding fast Fourier transform (FFT): (**a**) Fundamental optical Vernier effect, (**b–d**) first three harmonic orders. Dashed line: Upper envelope (shifted upward to be distinguishable from the internal ones). Red-orange lines: Internal envelopes. Green arrow: Frequency of the sensing interferometer (FPI$_1$). White-blue arrow: Frequency of the reference interferometer (FPI$_2$). The non-marked small peak in the FFTs corresponds to the frequency difference between the two interferometers.

For sensing applications, monitoring the wavelength shift of the upper envelope for higher harmonic orders seems to have a drawback: The visibility decreases with the order of the harmonics. Although, at first glance, this might seem to be a disadvantage, for practical applications, the problem is easily solved by alternatively relying on the internal envelopes, as represented in Figure 2. These envelopes are obtained by fitting groups of maxima in the harmonic spectrum. The maxima can be classified into groups of $i+1$ peaks, the same as the number of internal envelopes generated. The intersection between internal envelopes provides multiple points that can be used to track the wavelength shift, instead of using the upper envelope. Fitting the internal envelopes also reduces the effects of intensity fluctuations in different peaks, which might contribute to an error in the measurement of the wavelength shift. Additionally, the *FSR* of the internal envelope scales with the order of the harmonics, also visible in Figure 2, as the frequency of the reference interferometer (inverse of the *FSR*) increases harmonically with the order of the harmonics. Note that the detuning is the same for all the presented cases. The *FSR* of the internal envelope can be expressed as:

$$FSR_{internal\ envelope}^i = \left| \frac{(i+1)FSR_1 FSR_2^i}{FSR_1 - (i+1)FSR_2^i} \right| = (i+1)FSR_{envelope}^i, \tag{10}$$

where the internal envelope is larger by a factor of $i+1$ than the upper envelope (Equation (9)). If high finesse Fabry–Perot interferometers were used, the spectral dips would become narrower, which can be an advantage for tracking their position and tracing the envelopes. At the same time, the envelope properties of the effect would still be maintained.

As discussed in the previous section, the *M*-factor for the fundamental optical Vernier effect is obtained by dividing the *FSR* of the upper envelope by the *FSR* of the sensing interferometer. Although such calculations work for the fundamental effect, they are not correct for the harmonics. The result would be an *M*-factor independent of the order of the harmonics, since the *FSR* of the upper envelope is the same for every harmonic, as described by Equation (9). In fact, the *M*-factor does not depend on the upper envelope, but rather on the internal envelope, as we will later demonstrate. Therefore, the general expression for the *M*-factor as a function of the order of the harmonic is defined as:

$$M^i = \frac{FSR_{internal\ envelope}^i}{FSR_1} = \left| \frac{(i+1)FSR_2^i}{FSR_1 - (i+1)FSR_2^i} \right| = (i+1)M, \tag{11}$$

where the first interferometer (FPI₁) is assumed to be the sensing interferometer, and the second one (FPI₂) serves as a reference.

The *M*-factor for the fundamental optical Vernier effect is recovered for $i = 0$. In a situation where the OPL of the reference interferometer (FPI₂) is scaled up to generate harmonics of the Vernier effect, the magnification obtained scales up linearly with the order of the harmonic, for the same detuning. This detuning corresponds to the optical path difference between the actual reference interferometer and the closer situation of a perfect harmonic case (where $OPL_2 = (i+1)OPL_1$). When no detuning is considered, the magnification factor trends towards infinity, translating in a Vernier envelope with an infinite *FSR*. In practice, it corresponds to a useless situation where the Vernier envelope cannot be tracked and measured. In order to make the structure useful, one has to deliberately apply a detuning (Δ) to the reference interferometer OPL to slightly move away from the perfect harmonic situation.

For a harmonic of order *i*, the magnification increases $i+1$ times the value of the magnification for the fundamental optical Vernier effect. This means that the wavelength shift of the envelope also increases linearly with the order of the harmonic, allowing for the realization of sensors with a sensitivity enhanced by $i+1$ times.

In the fundamental optical Vernier effect, the maximum *M*-factor is limited in practical applications by the *FSR* of the upper envelope, where one period should stay within the wavelength range available from the detection system. In the case of the harmonic effect, the maximum *M*-factor is not directly limited by the *FSR* of the internal envelope, although it scales up with the order of the harmonic. Even if the period of the upper envelope stays out of the wavelength range available, one can still rely on the internal envelope intersections to monitor the wavelength shift, as discussed before.

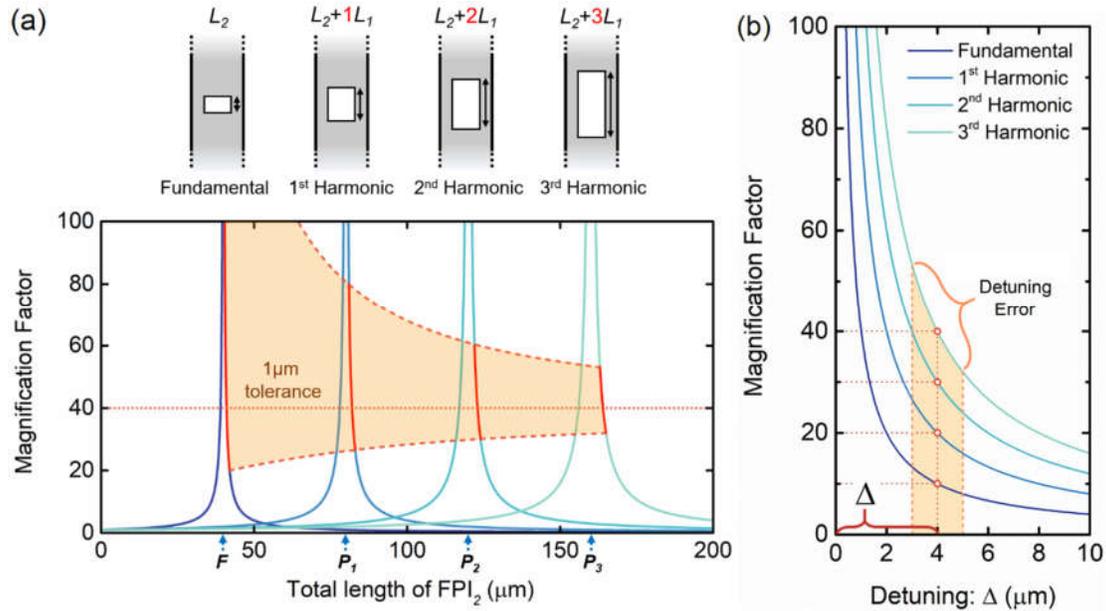

**Figure 3.** **(a)** Magnification factor as a function of the total length ( $L_2 + iL_1$ ) of the reference interferometer (FPI₂), for a fixed length ($L_1$) of the sensing interferometer (FPI₁), where $i$ corresponds to the order of the harmonic. The perfectly harmonic case is marked with $F$, $P_1$, $P_2$, and $P_3$, respectively, for the fundamental and the first three harmonic orders, where the $M$-factor is infinite. A deviation of 1 μm in the length of FPI₂ produces smaller variations in the $M$-factor for higher harmonic orders, as exhibited by the red line. **(b)** Magnification factor as a function of the detuning ($\Delta$) from a perfectly harmonic situation applied to the reference interferometer (FPI₂). For the same detuning, the magnification factor scales up linearly with the order of the harmonics as can be seen e.g. by the values at the red circles. Small detuning errors from multiple sources, such as fabrication tolerances, can modify the obtained magnification factor.

A different way to approach these concepts is represented in Figure 3b. Here, the magnification factor for the fundamental case and for the first three harmonic orders is plotted as a function of the detuning ($\Delta$) from the perfect harmonic case. One can see, e.g., by observing the magnification at the positions of the red circles that, for the same detuning (introduced on purpose to make the envelope measurable) the $M$-factor scales up linearly with the order of the harmonics. Even though it is not a perfect harmonic case, for a fixed detuning, the scaling properties of the effect (magnification factor, number of internal envelopes, frequency of the internal envelope) can still be seen as harmonic. Figure 3b also presents a 1 μm detuning error, showing how it can affect the final $M$-factor.

## 3. Results

In this section, an experimental demonstration on the production of harmonics of the Vernier effect is presented, demonstrating the improved sensitivity magnification properties previously discussed.

### 3.1. Experimental Setup

The experimental setup is illustrated in Figure 4a. The two Fabry–Perot interferometers (sensing and reference) are physically separated in a parallel configuration by means of a 50/50 fiber coupler. The sensing interferometer is connected to port 2, while the reference interferometer is connected to port 3. The input port 1 is connected to a supercontinuum laser source (Fianium WL-SC-400-2). The reflected signals from port 2 and 3 are combined and measured at port 4 with an optical spectrum analyzer (OSA ANDO AQ-6315A, resolution of 0.1 nm).

To perform the strain measurements, the sensor was glued to a fixed platform and to a translation stage with a resolution of 0.01 mm. The total length over which strain was applied, corresponding to the length between the fixed points, is 344 mm. All the experiments were carried out at room temperature (23 °C). The strain was applied to FPI$_1$, for all three cases of different reference FPIs. The strain measurements were done by applying strain up to 600 με, with steps of 87.2 με (0.03 mm). Only static measurements were performed in this experiment.

Figure 3a displays the *M*-factor curve, defined through Equation (11) as a function of the total length of the second interferometer, for a fixed dimension of the first interferometer. The *M*-factor trends toward infinity as the OPLs of the two interferometers become attuned, approaching a perfect harmonic situation. The points marked as *F*, *P$_1$*, *P$_2$*, and *P$_3$* correspond to the perfect harmonic case for the fundamental and the first three harmonic orders of the Vernier effect, respectively. This situation corresponds to an infinite *M*-factor. It is visible that the *M*-factor curve broadens for higher harmonic orders. First, this allows higher *M*-factors to be achieved more easily, and second, it allows the impact of small detuning errors to be reduced. There are different sources of detuning errors. Environmental effects, such as temperature changes or deformation/strain, would typically result in a percentage change in the interferometer length and would become more relevant for longer reference interferometers. Besides these environmental effects, errors, and tolerances in the fabrication process also contribute to detuning errors. From all these sources of error, strain, or deformation effects are negligible in our case, as the reference interferometer is considered stable since no strain is applied to it. The thermal expansion coefficient of silica is around $0.55 \times 10^{-6} K^{-1}$, which for a 5 °C temperature variation corresponds to a length variation of $2.75 \times 10^{-4}$ %. In practical terms, for a 100 μm-long cavity the length variation caused by this temperature variation is 0.275 nm, and for a 1 mm-long cavity that corresponds to a length variation of 0.275 μm. These variable parameters produce a detuning error, which is, in general, below the error imposed by the accuracy of the fabrication procedures (between 1 μm to a few micrometers). Therefore, the limiting factor here is the detuning error caused by the fabrication process, which is a fixed value dependent on the available fabrication technology. One can see by the red line of Figure 3a that the variation in the *M*-factor caused by a 1 μm fixed detuning error in the length of the second FPI is smaller for higher harmonic orders. Therefore, higher harmonic orders allow larger tolerances in sensor fabrication without compromising this parameter, for the same *M*-factor.

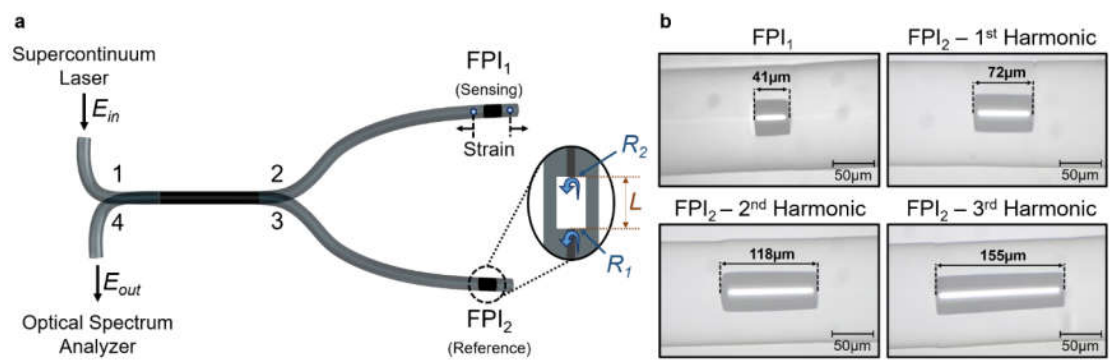

**Figure 4.** (**a**) Schematic illustration of the experimental setup. The sensing interferometer (FPI$_1$) and the reference interferometer (FPI$_2$) are separated by means of a 50/50 fiber coupler. A supercontinuum laser source is connected to the input and the reflected signal from the device is measured with an optical spectrum analyzer. Light is reflected at both interfaces of the capillary tube, with intensity reflectivities *R$_1$* and *R$_2$*. The length of the interferometer (*L*) is given by the length of the capillary tube. Strain is only applied to FPI$_1$. (**b**) Micrograph of the experimental fiber sensing interferometer (FPI$_1$) and the three different reference interferometers (FPI$_2$) used to excite the first three harmonic orders.

*3.2. Sensor Fabrication*

The Fabry–Perot interferometers used in the experiment were based on a section of a capillary tube spliced between two pieces of single-mode fiber (SMF28). The capillary tube, fabricated at Leibniz-IPHT, has an internal diameter of 60 μm and an outer diameter of 125 μm. First, the end of a single-mode fiber and a capillary tube were cleaved with a fiber cleaver and spliced together using a splicing machine (Fitel S177). The splice was performed in the manual mode of the fusion splicer, ensuring that the center of the electric arc was mainly applied to the single mode fiber, thus avoiding the collapse of the capillary tube. The following parameters were used: Two electric arc discharges with an arc power of 30 arb. units and arc duration of 400 ms. Then, the fiber was placed in the fiber cleaver and, with the help of a magnification lens, the capillary tube was cleaved with the desired length. At last, the cleaved end of the capillary tube was spliced to another piece of single-mode fiber following the same procedures as described before. Figure 4b shows a micrograph of the different Fabry–Perot interferometers fabricated. The sensing interferometer (FPI$_1$) is 41 μm long. Three reference interferometers (FPI$_2$) were fabricated to excite the first three harmonic orders of the Vernier effect $\left(OPL_2 = (i+1)OPL_1 + \Delta,\right.$ where $\Delta$ is the detuning$)$. An FPI$_2$ with a length of 72 μm (2 × 41 μm plus a detuning of -10 μm) was used to produce the first harmonic. The second harmonic was excited using an FPI$_2$ with a length of 118 μm (3 × 41 μm plus a detuning of -5 μm). At last, an FPI$_2$ with a length of 155 μm (4 × 41 μm plus a detuning of -9 μm) was used to produce the third harmonic.

Initially, the sensing interferometer (FPI$_1$), with an *FSR* of 23.52 nm, was characterized with regard to strain sensitivity, obtaining a value of (3.37 ± 0.02) pm/με. Then, three Fabry–Perot interferometers with different lengths were successively applied as the reference interferometer in order to respectively excite the first three harmonic orders of the Vernier effect.

Figure 5a–c depicts the experimental reflected intensity spectra for the first three harmonic orders, with different detunings. The appearance of the reflected intensity spectra is similar to the theoretical results, as predicted by Equation (4) and as shown in Figure 2. The number of internal envelopes scales up linearly with the order of the harmonics, providing intersection points suitable for monitoring the wavelength shift in sensing applications. The *FSR* of the upper envelope for the first three harmonic orders is 98.56, 222.80, and 107.77 nm, respectively. The *FSR* of the internal envelopes is given approximately by $i+1$ times the *FSR* of the upper envelope. With this, one can determine the *M*-factor for each harmonic through Equation (11). The *M*-factor obtained via the *FSR*s for the first three harmonic orders is 8.38, 28.42, and 18.33, respectively. Note that the *M*-factor is dependent on the detuning of the reference interferometer. As the *FSR* of the upper envelope is independent of the harmonic order, the larger the upper envelope, the less detuned the effect is and, therefore, the higher the *M*-factors achieved, as discussed before. For example, the second harmonic sensor is well-tuned, therefore it shows a larger envelope and a higher *M*-factor than the third harmonic, which is not so well-tuned (larger detuning).

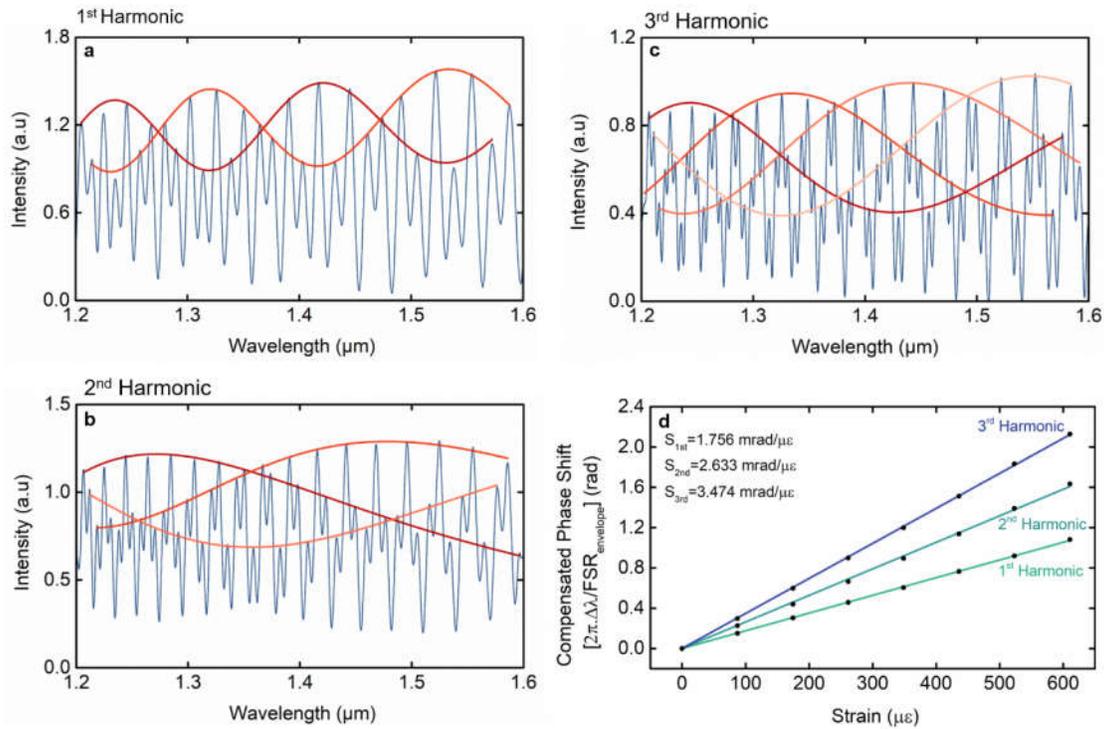

**Figure 5.** (**a–c**) Experimentally obtained reflected intensity spectra for the first three harmonic orders. Red-orange lines correspond to the internal envelopes. The number of internal envelopes increases linearly with the order of the harmonic, as expected theoretically. (**d**) The compensated phase shift $\left(2\pi \times \Delta\lambda/FSR_{envelope}\right)$ for the first three harmonic orders as a function of the strain applied. The compensated phase sensitivities to strain, given by the slope of the curves (*S*), increase linearly with the order of the harmonics, demonstrating the magnification enhancement predicted theoretically.

To further validate the properties of the effect, the first three harmonic orders were characterized in strain. An example of the experimental spectral shift of the harmonic Vernier spectrum can be found in Appendix D. Sensitivity values of (27.6 ± 0.1 pm/με), (93.4 ± 0.6) pm/με, and (59.6 ± 0.1) pm/με were achieved for the first, second, and third harmonics, respectively. Calculating the *M*-factor through the ratio between the sensitivity of the Vernier envelope and the sensitivity of the individual sensing interferometer (FPI₁) determined previously, one obtains 8.18, 27.7, and 17.7, respectively, for each harmonic. Both, the *M*-factors defined using the *FSR* of the internal envelopes (Equation (11)) and the *M*-factors defined using the sensitivities (Equation (7)), are approximately the same, with a maximum deviation of 3.5%. In other words, both definitions for the *M*-factor are equivalent.

In order to verify Equation (11), where the *M*-factor for each harmonic order is $i+1$ times larger than the *M*-factor for the fundamental optical Vernier effect, the ratio between the *M*-factor obtained via the sensitivities and that of the equivalent fundamental optical Vernier effect was calculated. This last factor is determined by the ratio between the *FSR* of each upper envelope, which is independent of the order of the harmonic, and the *FSR* of the individual sensing interferometer (FPI₁). Hence, the ratios obtained are 1.95, 2.93, and 3.86, respectively, for the first three harmonic orders. As observed, the ratio values are approximately increasing by factors of $i+1$, as predicted by Equation (11).

Furthermore, it is possible to make a fair comparison between the structures independent of the specific detuning value of the reference interferometers, and also to demonstrate more directly the linear enhancement of the *M*-factor with the order of the harmonics by using a compensated wavelength shift. The compensated wavelength shift takes into consideration the *FSR* of the upper envelope which, as we discussed before, is an indicator of the detuning of the structure. Therefore, the compensated wavelength shift $\left(\Delta\lambda/FSR_{envelope}\right)$ is independent of the tuning of these structures. It can also be transformed into a more meaningful value as $\left(2\pi \times \Delta\lambda/FSR_{envelope}\right)$, corresponding now to the envelope phase shift, independent of the detuning. A more comparable value of sensitivity, which quantifies how sensitive each interferometer is, can now be defined as $S = \left(\dfrac{2\pi \times \Delta\lambda}{FSR_{envelope}}\right)\!\Big/strain$, representing the compensated phase sensitivity to strain. Figure 4d represents the compensated wavelength shift for the first three harmonic orders. The compensated sensitivity for each structure, defined by the slope of the curves, increases linearly with the order of the harmonics, which is also in accordance with Equation (11).

A summary of the main values of the experimental results is displayed in Table 1. They can be organized into three groups. The first resumes the experimental results for strain sensitivity, where a compensated strain sensitivity is displayed for comparison purposes. The compensated strain sensitivity is a way to observe only the influence of the harmonics in the experimental sensitivity. It increases with the order of the harmonics, demonstrating the sensitivity enhancement property of the effect independently of the detuning. The second group shows the *M*-factors by the two different definitions (through Equations (11) and (7)). The values obtained are very similar, validating the use of both definitions. At last, the third group compares the *M*-factor for each harmonic with the *M*-factor for the equivalent fundamental optical Vernier effect. It shows the $i+1$ factor improvement in the *M*-factor with the order of the harmonic, as predicted by Equation (11).

**Table 1.** Overview of the experimental results for the first three harmonic orders. First group: Experimental results. Second group: M-factor by different definitions (Equations (7) and (11)) give the same results. Third group: M-factor for each harmonic compared with the M-factor for the fundamental optical Vernier effect. It shows the $i+1$ improvement factor with the order of the harmonic.

| | Experimental Strain Sensitivity (pm/µε) | Compensated Phase Sensitivity (mrad/ µε) | *M*-Factor by $FSR_{envelope}$ Equation (11) | *M*-Factor by Sensitivities Equation (7) | *M-factor by FSR / M-factor by Fundamental Vernier Effect* |
|---|---|---|---|---|---|
| 1st Harmonic | 27.6 | 1.765 | 8.38 | 8.18 | 1.95 **(2)** |
| 2nd Harmonic | 93.4 | 2.633 | 28.41 | 27.70 | 2.93 **(3)** |
| 3rd Harmonic | 59.6 | 3.474 | 18.32 | 17.70 | 3.86 **(4)** |

## 4. Discussion

The fundamental optical Vernier effect can be used to achieve high sensitivity magnification values. However, these magnifications are limited by the wavelength range available in the detection system. The generation of optical harmonics of the Vernier effect is an effective tool to further increase, by several fold, the magnification values and to achieve higher resolutions without compromising the signal detection and monitoring. The presence of internal envelopes, which are different from the upper envelope, typically monitored in the fundamental optical Vernier effect, provides intersection points better suited to tracking the wavelength shift used for sensing.

In practical applications, fine-tuning of the interferometer optical path length can be quite complicated and, in most cases, technically challenging. Therefore, going for higher-order harmonics allows not only better fabrication tolerances, but also new ways of dimensioning interferometric structures and new sensing configurations.

From a theoretical point of view, the equations allow the use of harmonic orders that can go up to infinity. However, from an experimental perspective, the interference peaks get narrower, and after a certain harmonic order, their *FSR* may stay below the resolution of the detection system. Therefore, the peaks become indistinguishable, and internal envelopes fitting is no longer accessible. Moreover, as the modulation contrast of the upper envelope decreases with the order of the harmonics, the visibility of the modulation will deteriorate due to signal noise. Therefore, the limitation in terms of the maximum harmonic order achievable depends both on the specific application and on the available detection system.

The possibility of discriminating between multiple parameters, like strain and temperature, should be explored in the future. Apart from the envelopes, the Fabry–Perot-like response is still present in the spectrum. Its response, together with the envelope, could be used with a matrix method to discriminate between different measurands or, for instance, compensate fluctuations in temperature.

Another interesting point to discuss is the possibility of performing dynamic measurements with the Vernier effect. One drawback of the method is the need to measure a certain wavelength range, which is in general large, in order to be able to trace and track the envelope. Indeed, such requirements impose a limitation for dynamic measurements, especially if an OSA or other slow speed detection system is used. One way to tackle this problem is to design the structure to work in the wavelength range of a commercially available fast optical interrogator, together with specific software to extract the envelope and measure the wavelength shift. A different solution would be to study and explore the possibility of developing a detection system, e.g., with a wavelength tunable laser source that could rapidly track the envelope shift without the need to measure a full spectrum and avoid the use of an OSA.

In sum, the use of optical harmonics of the Vernier effect enables the exploitation of a new generation of sensors, capable of fulfilling the sensitivity and resolution requirements for state-of-the-art applications in areas like medicine, biology, and chemistry. With this setup, one can boost the performance of conventional interferometric sensors to unprecedented values while making their fabrication flexible and adaptable to a specific application. As an example of future development, one could achieve sub-nanostrain resolution with a simple fiber geometry, such as a silica tube.

**Author Contributions:** Conceptualization, A.G. and M.F.; methodology, A.G. and M.F.; validation, A.G., M.F., and H.B.; formal analysis, A.G., M.F., and H.B.; investigation, A.G. and M.F.; resources, J.B., J.K., and M.R.; data curation, A.G.; writing—original draft preparation, A.G.; writing—review and editing, M.F., J.B., J.K., M.R., H.B., and O.F.; visualization, A.G. and M.F.; supervision, H.B. and O.F.; project administration, M.R. and O.F.; funding acquisition, M.R.;

**Funding:** This work was partially financed by the ERDF – European Regional Development Fund through the Operational Programme for Competitiveness and Internationalization - COMPETE 2020 Programme and by National Funds through the Portuguese funding agency, FCT - Fundação para a Ciência e a Tecnologia within the ENDOR project (POCI-01-0145-FEDER-029724). The work was also supported by FEDER funds through the COMPETE 2020 Programme and National Funds through FCT under the project UID/CTM/50025/2013. André D. Gomes is grateful for the PhD scholarship funded by FCT (SFRH/BD/129428/2017). Marta S. Ferreira is also grateful for the research fellowship no. SFRH/BPD/124549/2016. The publication of this article was funded by the Open Access Fund of the Leibniz Association.

**Acknowledgments:** The authors would like to acknowledge Katrin Uhlig for the contribution to the design of Figure 1.

**Conflicts of Interest:** The authors declare no conflict of interest.

## Appendix A. Derivation of the reflected light intensity for the fundamental optical Vernier effect

Let us consider a parallel configuration, where two Fabry–Perot interferometers (FPIs) are physically separated by using a 3dB fiber coupler [4]. For simplification, the two interferometers are assumed to have identical interfaces, with an intensity reflection coefficient $R_1$ for the first interface and $R_2$ for the second one.

Although the following theoretical considerations are valid for any FPI structure, we will assume that each FPI is generated by a section of a silica tube between single-mode fibers. In this case, all interfaces provide a silica/air Fresnel reflection. Note that the reflection coefficient due to a silica/air Fresnel reflection is small (around 3.3% at 1550 nm), and hence, only one reflection at each interface was considered, as a two-wave approximation.

In this configuration, light is injected at port 1 and split between the two arms (port 2 and 3) with equal intensity. The light reflected by the system is collected at port 4. The electric field of the input light, $E_{in}(\lambda)$, propagating in the structure, will be reflected at different points. In both interferometers, the electric field of light reflected at the interface 1 of the interferometer is given by:

$$E_{R1}(\lambda) = \sqrt{R_1}\,\frac{E_{in}(\lambda)}{\sqrt{2}}, \tag{A1}$$

while the electric field of the transmitted light at the same interface is expressed as:

$$E_{T1}(\lambda) = \sqrt{(1-A_1)}\sqrt{(1-R_1)}\,\frac{E_{in}(\lambda)}{\sqrt{2}}, \tag{A2}$$

where $A_1$ represents the transmission losses through interface 1, related to mode mismatch and surface imperfections.

Light transmitted at the interface 1 $\left(E_{T1}(\lambda)\right)$ will then travel through the FPI, being partially reflected and transmitted at the interface 2. The electric field of the light reflected at the interface 2 of the first interferometer is expressed as:

$$E_{R2}^1(\lambda) = \sqrt{(1-A_1)}\sqrt{(1-R_1)}\exp(-\alpha L_1)\sqrt{R_2}\,\frac{E_{in}(\lambda)}{\sqrt{2}}\exp\left[-j\left(2\pi n_1 L_1/\lambda - \pi\right)\right], \tag{A3}$$

where $\exp(-\alpha L_1)$ represents the propagation losses in the first interferometer, $2\pi n_1 L_1/\lambda - \pi$ is the phase accumulated in the propagation up to interface 2, with a reflection phase of pi, $\lambda$ is the vacuum wavelength of the input light, $n_1$ and $L_1$ are the effective refractive index and the length of the first interferometer. This reflected light $\left(E_{R2}^1(\lambda)\right)$ will be propagated back into the structure and get partially transmitted at the interface 1 towards the output, interfering with the light initially reflected at that interface as described by Equation (A1). Therefore, the electric field of the light coming from the first interferometer is given by:

$$E_{FPI1}(\lambda) = \frac{E_{in}(\lambda)}{\sqrt{2}}\left\{\sqrt{R_1} + (1-A_1)(1-R_1)\exp\left(-\alpha 2L_1\right)\sqrt{R_2}\exp\left[-j\left(4\pi n_1 L_1/\lambda - \pi\right)\right]\right\}. \tag{A4}$$

At the interface 2, the transmitted light leaves the interferometer, and therefore no longer contributes to the system.

The same analysis can be performed for the second interferometer, where the electric field of the light coming from the second interferometer can be expressed, in a similar form as in Equation (A4), as:

$$E_{FPI2}(\lambda) = \frac{E_{in}(\lambda)}{\sqrt{2}}\left\{\sqrt{R_1} + (1-A_1)(1-R_1)\exp\left(-\alpha 2L_2\right)\sqrt{R_2}\exp\left[-j\left(4\pi n_2 L_2/\lambda - \pi\right)\right]\right\}, \tag{A5}$$

where $n_2$ and $L_2$ are the effective refractive index and length of the second interferometer.

The electric field of the light coming from the first interferometer, expressed by Equation (A4), is redirected from port 2 of the couple into the output. It can be simplified as:

$$E_R^{port2}(\lambda) = \frac{E_{in}(\lambda)}{\sqrt{2}}\left\{A + B\exp\left[-j\left(4\pi n_1 L_1/\lambda - \pi\right)\right]\right\}, \tag{A6}$$

with $A$ and $B$ given by:

$$A = \sqrt{R_1}, \tag{A7}$$

$$B = (1-A_1)(1-R_1)\exp(-2\alpha L_1)\sqrt{R_2}. \tag{A8}$$

In the same way, the reflected electric field coming from port 3 towards the output is expressed as:

$$E_R^{port3}(\lambda) = \frac{E_{in}(\lambda)}{\sqrt{2}} \left\{ A + C \exp\left[-j\left(4\pi n_2 L_2 / \lambda - \pi\right)\right] \right\},\qquad(A9)$$

where $C$ corresponds to:

$$C = (1 - A_1)(1 - R_1)\exp(-2\alpha L_2)\sqrt{R_2}.\qquad(A10)$$

If no propagation losses are considered as a simplification, the coefficients described by equations (A8) and (A10) are the same $(B = C)$. With this, one can express the total electric field leaving the output at port 4 $\left( E_R^{port2} + E_R^{port3} \right)$ as:

$$E_{out}(\lambda) = \sqrt{2}A E_{in}(\lambda) + B\frac{E_{in}(\lambda)}{\sqrt{2}} \left\{ \exp\left[-j\left(4\pi n_1 L_1 / \lambda - \pi\right)\right] + \exp\left[-j\left(4\pi n_2 L_2 / \lambda - \pi\right)\right] \right\},\quad(A11)$$

where $B$ is now defined as:

$$B = (1 - A_1)(1 - R_1)\sqrt{R_2},\qquad(A12)$$

The output light intensity, $I_{out}(\lambda)$, normalized to the incident light, can now be calculated by:

$$I_{out}(\lambda) = \left| \frac{E_{out}(\lambda)}{E_{in}(\lambda)} \right|^2 = \frac{E_{out}(\lambda)E_{out}^*(\lambda)}{E_{in}^2(\lambda)},\qquad(A13)$$

where $E_{out}^*(\lambda)$ is the complex conjugated of $E_{out}(\lambda)$. By substituting Equation (A11) in Equation (A13), after some algebraic manipulation the expression for the reflected light intensity measured at the output is:

$$I_{out}(\lambda) = I_0 - 2AB\left[\cos\left(4\pi n_1 L_1 / \lambda\right) + \cos\left(4\pi n_2 L_2 / \lambda\right)\right] + B^2 \cos\left[4\pi\left(n_1 L_1 - n_2 L_2\right)/\lambda\right],\qquad(A14)$$

where $I_0 = 2A^2 + B^2$.

## Appendix B. Derivation of the free spectral range (*FSR*) of the envelope and magnification factor (*M*-factor) for the fundamental Vernier effect

Let us consider the overlap of two Fabry–Perot interferometers with slightly shifted interferometric frequencies, whose individual responses are represented in Figure A1. The interferometer 1, corresponding to the orange curve, is slightly smaller than interferometer 2, given by the blue curve.

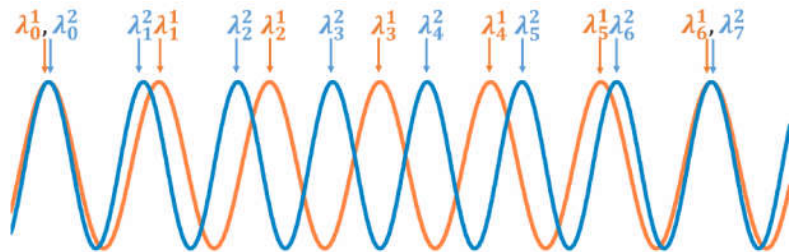

**Figure A1**. Diagram of the response of two Fabry–Perot interferometers (1 and 2). The wavelengths of the different peaks are labeled as $\lambda_k^m$, where $m = 1, 2$ is the number of the interferometer and $k$ is the number of the peak.

The wavelengths of the different peaks are labeled as $\lambda_k^m$, where $m = 1, 2$ is the number of the interferometer and $k$ is the number of the peak. From Figure A1, both interferometers are in phase at a wavelength $\lambda_0^m$. The wavelength of the maximum "$k$" can be described using the free spectral range ($FSR$) of the interferometer as:

$$\lambda_k^1 = \lambda_0 + kFSR_1,$$ (A15)

for the interferometer 1, and similarly as:

$$\lambda_k^2 = \lambda_0 + kFSR_2,$$ (A16)

for the interferometer 2. At a certain wavelength, both interferometers will be once again in phase. In this case, from Figure A1 one can see that both interferometers are again in phase when:

$$\lambda_k^1 = \lambda_{k+1}^2.$$ (A17)

Introducing Equations (A15) and (A16) into Equation (A17), the following relationship is obtained:

$$kFSR_1 = (k+1)FSR_2.$$ (A18)

Therefore, one can express "$k$" as a function of the $FSR$s of both interferometers in the form of:

$$k = \frac{FSR_2}{FSR_1 - FSR_2}.$$ (A19)

In the fundamental optical Vernier effect, the $FSR$ of the envelope corresponds to the wavelength distance between two consecutive situations where both interferometers are in phase. Hence, the $FSR$ of the envelope can be given by:

$$FSR_{envelope} = \lambda_k^1 - \lambda_0 = kFSR_1$$ (A20)

By substituting Equation (A19) in Equation (A20), the final expression for the $FSR$ of the envelope is:

$$FSR_{envelope} = \frac{FSR_2 FSR_1}{FSR_1 - FSR_2}.$$ (A21)

In a Fabry–Perot interferometer, the $FSR$ is usually defined as [32]:

$$FSR = \frac{\lambda_1 \lambda_2}{2nL},$$ (A22)

where $\lambda_1$ and $\lambda_2$ are the wavelengths of two consecutive maxima (or minima), $n$ is the effective refractive index of the interferometer, and $L$ is the length of the interferometer. With this, one can define the $FSR$ of the envelope in a different way as:

$$FSR_{envelope} = \left| \frac{\lambda_0^1 \lambda_k^1}{2(n_1 L_1 - n_2 L_2)} \right|.$$ (A23)

The magnification factor ($M$) is defined as the ratio between the $FSR$ of the envelope and the $FSR$ of the individual sensing interferometer. If the interferometer 1 is the sensing interferometer, the $M$-factor is defined as:

$$M = \frac{FSR_{envelope}}{FSR_1} = \left| \frac{FSR_2}{FSR_1 - FSR_2} \right|,$$ (A24)

which is the same as the index "$k$" defined in Equation (A19). If we substitute definitions described by equations (A22) and (A23) for the $FSR$s, with the rough approximation of $\lambda_1^1 \lambda_2^1 = \lambda_0^1 \lambda_k^1$, the magnification factor can also be expressed as:

$$M = \left| \frac{n_1 L_1}{n_1 L_1 - n_2 L_2} \right|.$$ (A25)

### Appendix C. Derivation of the free spectral range (*FSR*) of the envelope for the first optical Vernier effect harmonic and generalization for any harmonic order

Let us consider the same two Fabry–Perot interferometers as in Appendix B, but now with the optical path length of interferometer 2 increased by one-time the optical path length of interferometer 1, $OPL_2 = n_2L_2 + in_1L_1$ with $i = 1$, in order to produce the first harmonic. The individual responses of the two interferometers are represented in Figure A2 (orange curve: interferometer 1; blue curve: interferometer 2).

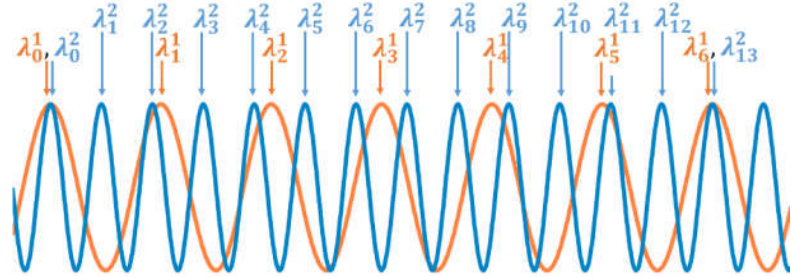

**Figure A2**. Diagram of the response of two Fabry–Perot interferometers (1 and 2), with the optical path of 2 increased by one-time the optical path of 1 $\left( OPL_2 = n_2L_2 + 1n_1L_1 \right)$.

Due to this change in the optical path length of the second interferometer, the *FSR* of interferometer 2 is now defined as:

$$FSR_2^{i=1} = \frac{\lambda_1\lambda_2}{2\left(n_2L_2 + 1n_1L_1\right)}. \tag{A26}$$

The same analysis as in Appendix B can be performed. From Figure A2, both interferometers are in phase at a wavelength $\lambda_0^m$. The wavelength of the maximum "$k$" can be described using the free spectral range (*FSR*) of the interferometer as:

$$\lambda_k^1 = \lambda_0 + kFSR_1, \tag{A27}$$

for the interferometer 1, and similarly as:

$$\lambda_k^2 = \lambda_0 + kFSR_2^{i=1}, \tag{A28}$$

for the interferometer 2. At a certain wavelength, both interferometers will be once again in phase. In this case, from Figure A2 one can see that both interferometers are again in phase when:

$$\lambda_k^1 = \lambda_{2k+1}^2. \tag{A29}$$

Introducing equations (A27) and (A28) into Equation (A29), the following relationship is obtained:

$$kFSR_1 = \left(2k+1\right)FSR_2^{i=1}. \tag{A30}$$

Therefore, one can express "$k$" as a function of the *FSR*s of both interferometers in the form of:

$$k = \frac{FSR_2^{i=1}}{FSR_1 - 2FSR_2^{i=1}}. \tag{A31}$$

In the Vernier effect, the *FSR* of the envelope corresponds to the wavelength distance between two consecutive situations where both interferometers are in phase. Hence, the *FSR* of the envelope can be given by:

$$FSR_{envelope} = \lambda_k^1 - \lambda_0 = kFSR_1 \tag{A32}$$

By substituting Equation (A31) in Equation (A32), the final expression for the *FSR* of the envelope is:

$$FSR_{envelope} = \frac{FSR_2^{i=1}FSR_1}{FSR_1 - 2FSR_2^{i=1}}. \tag{A33}$$

In general, for a harmonic of order $i$, the optical path of interferometer 2 is increased by *i*-times the optical path of interferometer 1 $\left(OPL_2 = n_2L_2 + in_1L_1\right)$. Therefore, Equation (A26) for the *FSR* of interferometer 2 can be expressed in a general form as:

$$FSR_2^i = \frac{\lambda_1\lambda_2}{2\left(n_2L_2 + in_1L_1\right)}, \quad i = 0,1,2\dots \tag{A34}$$

Starting from an initial in-phase situation, both interferometers will be again in phase when:

$$\lambda_k^1 = \lambda_{(i+1)k+1}^2, \tag{A35}$$

being $i$ the order of the harmonic. Hence, one obtains a general expression for Equation (A30) as:

$$kFSR_1 = \left[\left(i+1\right)k+1\right]FSR_2^i, \tag{A36}$$

where "$k$" is now defined as:

$$k = \frac{FSR_2^i}{FSR_1 - \left(i+1\right)FSR_2^i}, \tag{A37}$$

and, therefore, the general expression for the *FSR* of the envelope as a function of the order of the harmonic is:

$$FSR_{envelope} = \frac{FSR_2^i FSR_1}{FSR_1 - \left(i+1\right)FSR_2^i}. \tag{A38}$$

## Appendix D. Example of the spectral shift

An example of the experimental wavelength shift of the harmonic Vernier effect is depicted in Figure A3. The spectra correspond to the second harmonic of the Vernier effect, represented in Figure 4b. The intersection between internal envelopes can be monitored as a function of the applied strain. In this case, tracking the intersection around 1400 nm, one can observe a shift of the intersection position towards longer wavelengths with increasing applied strain.

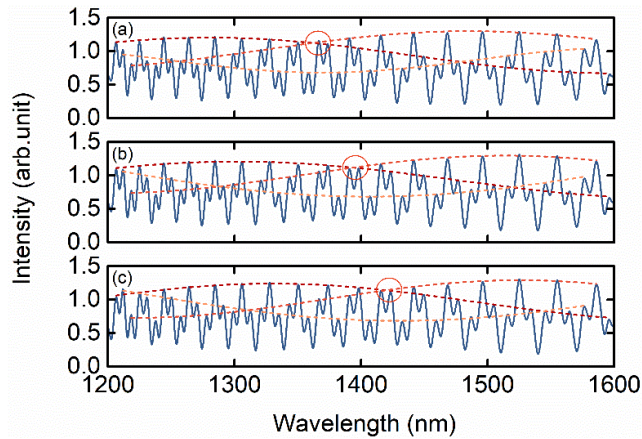

**Figure A3**. Demonstration of the experimental wavelength shift of the harmonic Vernier spectrum for three different strain values: (**a**) 0 με, (**b**) 348.8 με (**c**) 610.5 με. One of the multiple intersections between internal envelopes is marked with a red circle. The existence of a wavelength shift of the envelopes towards longer wavelengths is clearly visible as strain is applied.